\documentclass[twoside,11pt]{article}
\usepackage[]{graphicx}
\usepackage{float}
\usepackage{amsmath}
\usepackage{amssymb}
\usepackage{adjustbox}
\usepackage{tikzcd}
\usepackage{standalone}
\usepackage[T1]{fontenc}
\usepackage[utf8]{inputenc}
\usepackage[margin=1in]{geometry}
\usepackage{relsize}
\usepackage{amsthm}
\usepackage{sectsty}
\usepackage[all]{nowidow}

\usepackage[charter]{mathdesign}

\sectionfont{\fontsize{14}{15}\selectfont}
\subsectionfont{\fontsize{12}{15}\selectfont}

\newtheorem{theorem}{Theorem}[]
\newtheorem{lemma}{Lemma}[]
\newtheorem{property}{Property}[]
\newtheorem{corollary}{Corollary}[]

\begin{document}
\title{\Large A near-linear time minimum Steiner cut algorithm for planar graphs}
\author{Stephen Jue, Philip N. Klein}
\date{\today	}
\maketitle

\begin{abstract}
We consider the \textsc{Minimum Steiner Cut} problem on undirected planar
graphs with non-negative edge weights. This problem involves finding the
minimum cut of the graph that separates a specified subset $X$ of vertices
(\emph{terminals}) into two parts. This problem is of theoretical interest
because it generalizes two classical optimization problems, \textsc{Minimum
}$s$-$t$\textsc{ Cut} and \textsc{Minimum Cut}, and of practical importance
because of its application to computing a lower bound for \textsc{Steiner
(Subset) TSP}. Our algorithm has running time $O(n\log{n}\log{k})$ where $k$ is
the number of terminals.

\end{abstract}
\section{Introduction}
The study of the \textsc{Minimum }$s$-$t$\textsc{ Cut} and \textsc{Minimum Cut}
problems on graphs has a rich history. In 1956, Ford and Fulkerson gave an
algorithm for the \textsc{Maximum }$s$-$t$\textsc{ Flow} problem and proved the
problem's duality to \textsc{Minimum }$s$-$t$\textsc{ Cut} \cite{ford}. The 
\textsc{Minimum Cut} problem was initially considered to be a harder variant of
the $s$-$t$ formulation, but further algorithmic improvements showed that they
have comparable difficulty. Indeed, for planar undirected graphs, the fastest
algorithm for \textsc{Minimum }$s$-$t$\textsc{ Cut} is $O(n\log{\log{n}})$
\cite{italiano} and the fastest algorithm for \textsc{Minimum Cut} is also
$O(n\log{\log{n}})$ \cite{lacki}.

The \textsc{Minimum Steiner Cut} problem is as follows. Given an undirected
edge-weighted graph $G$ and a subset of $k$ vertices $X \subseteq V$ called
terminals, find a \emph{terminal-separating} cut that minimizes the weight of
edges whose ends are separated by the cut. A cut $W \subseteq V$ is
\emph{terminal-separating} if there is at least one terminal in $W$ and at
least one terminal in $V\setminus W$. This problem generalizes both the 
\textsc{Minimum }$s$-$t$\textsc{ Cut} problem ($X = \{s, t\}$) and the
\textsc{Minimum Cut} problem ($X = V$).

\subsection{Our result}
We give a near-linear-time algorithm for \textsc{Minimum Steiner Cut}
in planar graphs.
\begin{theorem}\label{algorithm}
	Let $G = (V, E)$ be an undirected planar graph with $n$ vertices and edge
weights. Let $X \subseteq V$ be a set of $k$ terminals. The minimum Steiner cut
of $G$ with respect to $X$ can be found in time $O(n\log{n}\log{k})$.
\end{theorem}

A trivial algorithm repeats \textsc{Minimum }$s$-$t$\textsc{ Cut} on any terminal and the $k-1$ other terminals, giving a $O(n(\log{\log{n}})k)$ runtime. Our algorithm improves on this bound when $k > \log{n}$. Our algorithm follows the structure of a near-linear-time algorithm for
\textsc{Minimum Cut} in planar graphs due to Chalermsook, Fakcharoenphol and
Nanongkai~\cite{chalermsook}.
Their algorithm uses a divide-and-conquer strategy that is based on
finding a balanced cycle separator consisting of two shortest paths.
At its core, it uses calls to a subroutine for minimum $st$-cut.

The algorithm of Chalermsook \emph{et al.} (hereafter CFN) 
depends on the minimum $st$-cut for carefully selected vertices $s$
and $t$ being a candidate solution for \textsc{Minimum Cut}.  The CFN
algorithm cannot be
straightforwardly generalized to \textsc{Minimum Steiner Cut} because
that minimum $st$-cut might not be a terminal-separating cut.
However, we show that an appropriate terminal-separating cut can be
found by an algorithm that combines (1) some ideas of Reif's
\textsc{Minimum Cut} algorithm for planar graphs~\cite{reif} with (2)
an orientation of nontree edges with respect to a shortest-path tree.
The latter idea draws on work by Rao~\cite{Rao92} and subsequent work
by Park and Phillips~\cite{ParkPhillips}; however, their algorithms
were $\Omega(n^2)$ so our algorithm must use the structure differently.

Why do we restrict ourselves to planar graphs? Because the 
algorithms for minimum cut and minimum $s$-$t$ cut / maximum 
$s$-$t$ flow in general graphs have quadratic runtime. Hao and 
Orlin's minimum cut algorithm which runs in time 
$O(nm\log{(n^2/m)})$ could likely be modified to find the 
minimum Steiner cut of a general graph in the same amount of 
time \cite{hao}. Orlin's more recent maximum $s$-$t$ flow 
algorithm runs in time $O(nm)$ or $O(n^2 / \log{n})$ in the 
case when $m = O(n)$, which holds for planar graphs 
\cite{orlin}. These runtimes suggest a quadratic runtime for 
minimum Steiner cut using existing techniques. For applications 
such as computing a lower bound for Steiner TSP, faster 
runtimes are needed.

\subsection{Related work}

When it was published, the CFN algorithm had a running time of
$O(n \log^2 n)$.  As noted by Mozes, Nikolaev, Nussbaum, and
Weimann~\cite{mozes2018minimum}, this can be improved to
$O(n \log n \log \log n)$ by using the $O(n \log \log n)$
minimum-$st$-cut algorithm of Italiano, Nussbaum, Sankowski, and
Wulff-Nilsen~\cite{italiano}.

Borradaile \emph{et al.}~\cite{borradaile} gave an $O(n\log^3{n})$ algorithm to process a bounded-genus graph and
construct a data structure that supports constant-time $s$-$t$ minimum cut
queries.  By using the algorithm of Borradaile et al., one can
therefore find a minimum terminal-separating cut in $O(n\log^3{n})$ time.
Thus our algorithm gives a logarithmic-factor improvement for the
problem of finding a minimum terminal-separating cut.  (Our algorithm
is also much simpler than the very sophisticated algorithm of
Borradaile et al., and in fact we are in the process of implementing
our algorithm.)

For general \emph{unweighted} graphs, Cole and Hariharan~\cite{cole} gave a
 $O(C^3 n \log n + m)$ algorithm for computing the minimum 
 Steiner cut, if the size of that cut is $C$.   Bhalgat \emph{et al.}~\cite{bhalgat} later 
 improved this to $\tilde{O}(C^2 n + m)$.


\subsection{Application to obtaining lower bounds for Steiner TSP in planar
graphs} 
Our motivation to work on this problem comes from the \emph{Steiner
traveling salesman problem (TSP) in planar graphs}.  In this problem, one is
given the same input---an edge-weighted planar graph and a subset of
terminals, and the goal is to find a minimum-weight closed walk
visiting all terminals.\footnote{In studying TSP in general graphs
  (the metric TSP), one need not consider the Steiner variant; one can
always consider the metric induced on the terminals.  However, when
the graph is from a restricted family, e.g. the family of planar
graphs, this reduction is not applicable because the induced metric
likely lies outside the family.}

For standard (non-Steiner) TSP, the
well-known \emph{Held-Karp lower bound}, which is considered a good
lower bound in practice, is equivalent to solving a huge structured
linear program.  There has been much work on fast approximation
schemes for computing the Held-Karp lower bound; recently Chekuri
and Quanrud~\cite{chekuri2017approximating} showed that this could be done in
nearly linear time, by using the packing-covering framework,
a.k.a. multiplicative weights (see,
e.g.,~\cite{plotkin1995fast,KolliopoulosY05,AroraHK12}).  Their
algorithm (like at least one previous 
approximation scheme for Held-Karp~\cite{plotkin1995fast}) relies on
repeatedly finding a minimum cut in a graph whose weights change in
every iteration.  

In previous experimental work on an implementation~\cite{BeckerFKM17} of
(non-Steiner)
TSP in planar graphs~\cite{Klein08}, we computed lower bounds that are
slightly weaker than Held-Karp by optimizing over the \emph{subtour
  elimination polytope}.\footnote{This polytope is defined by $2^n$
  constraints.  The Held-Karp lower bound is equivalent to optimizing
  over the same polytope intersected with that defined by an additional
  $n$ constraints.}  We used an approximation algorithm that also relies on
repeatedly finding a minimum cut in a graph whose weights change in
every iteration.  For this purpose, we used the CFN algorithm.\footnote{This
was essential because the available implementations of Held-Karp required
 $\Omega(n^2)$ space and time, and we were testing on huge
sparse graphs.}  Using this implementation, we were able to show our
TSP implementation found near-optimal solutions in huge planar graphs.

We are now implementing an approximation scheme~\cite{Klein06} for Steiner TSP
in
edge-weighted planar graphs.\footnote{This problem models, e.g.,
  optimizing deliveries in a road map.}
For Steiner TSP, the Held-Karp lower bound and subtour elimination
lower bound are not valid.  However, one can consider a variant of
these linear programs in which the constraints that give rise to the
minimum-cut subproblem are replaced with constraints that give rise to
minimum-Steiner-cut subproblems.  Our plan is to use our new
minimum-Steiner-cut algorithm in an approximation scheme to compute
lower bounds on Steiner TSP and use these to evaluate the quality of
solutions obtained by our Steiner TSP implementation.

Note that the methods of previous approximation
schemes~\cite{plotkin1995fast,chekuri2017approximating} are not applicable
to computing a lower bound on Steiner TSP.  Even the original
minimum-spanning-tree method of Held and Karp~\cite{held1970traveling}
is not applicable.

\section{Preliminaries}
Let $G = (V, E)$ be an undirected planar graph.
It is useful to consider each (undirected) edge $e = uv$ as corresponding to a pair
of \emph{darts} pointing
in opposite directions, $u \rightarrow v$ and $v \rightarrow u$. For a dart $u
\rightarrow v$, $u$ is the \emph{tail} and $v$ is the \emph{head}. We denote
the set of darts as $D = E \times \{1, -1\}$.

\subsection{Cuts} Given an undirected graph $G=(V,E)$ and a proper
nonempty subset $S$ of its vertices, $\delta(S)$ denotes the set of
edges $uv$ such that $u$ is in $S$ and $v$ is not.  Such a set is
called a \emph{cut}.  We call it a \emph{simple} cut (sometimes called
a \emph{bond}) if $S$ and $V-S$ each induces a connected subgraph of
$G$.  Given a graph with edge-weights, the \textsc{Minimum Cut}
problem is to compute a cut whose total weight is minimum.  In the
\textsc{Minimum }$s$-$t$\textsc{ Cut} problem, one is given in
addition vertices $s$ and $t$, and one seeks a minimum-weight cut
$\delta(S)$ such that $s\in S$ and $t\not\in S$.  In the
\textsc{Minimum Steiner Cut} problem, one is given a subset $X$ of vertices (the \emph{terminals}), and one seeks a minimum-weight cut
$\delta(S)$ such that at least one terminal is in $S$ and at least one
is not.  Assume without loss of generality that the input graph $G$ is
connected.  In this case, for each of the above cut problems, there is
always a solution that is a \emph{simple} cut.

\subsection{Cycles and orientations}

An \emph{orientation} of a graph is a function $\lambda: E \rightarrow D$ that
maps each edge to one of its darts. We sometimes consider an orientation
$\lambda: E' \rightarrow D$ of a subset of edges $E' \subseteq E$ of a graph.
An \emph{oriented cycle} $(C, \lambda_C)$ is a cycle with an orientation on its edges such that,
for each consecutive  pair of edges $e_1$, $e_2$, the head of $\lambda_C(e_1)$ is the tail of
$\lambda_C(e_2)$. The orientation of a cycle can be clockwise or counterclockwise.
We say that an oriented cycle $(C, \lambda_C)$ \emph{agrees with} an
orientation $\lambda$ if $\lambda_C(e) = \lambda(e)$ for all edges $e$ in the domain of $\lambda$.
Two cycles $C_1$ and $C_2$ can be
\emph{combined} using their symmetric difference $C_1 \triangle C_2 = (C_1
\setminus C_2) \cup (C_2 \setminus C_1)$. We say that two oriented cycles
\emph{combine simply} if they combine to form a simple cycle that agrees with
their orientations.

\subsection{Planar graph duality} Let $G$ be a planar embedded graph.
The \emph{dual} of $G$ is another planar embedded
graph $G^*$ whose vertices correspond to faces of $G$ and vice versa. We refer
to $G$ as the \emph{primal}. Two vertices of $G^*$ are joined by an edge if
their corresponding faces are adjacent in $G$. Hence, every edge $e$ in $G$
corresponds to an edge $e^*$ in $G^*$, every vertex $v$ in $G$ corresponds to a
face $v^*$ in $G^*$, and every face $f$ in $G$ corresponds to a vertex $f^*$ in
$G^*$. For a dart $d$ of $G$, the corresponding dart of $G^*$ points from $d$'s left face (when $d$ is oriented upwards) to $d$'s right face.

\subsection{Interdigitating spanning trees} For every spanning tree $T$ of a
planar graph $G$, the set of edges of $G^*$ whose edges are not in $T$
form a spanning tree $T^*$ of $G^*$.  This tree is called the \emph{interdigitating tree}
with respect to $T$ (and vice versa). The \emph{fundamental cycle of $e$ in $T$} is the unique cycle consisting of $e$ together with the path in $T$ between its endpoints. Let $(T^*, \lambda)$ be the interdigitating tree with its edges oriented rootwise. This orientation corresponds to a counterclockwise orientation to each non-tree edge $e$ and its fundamental cycle in the primal graph $G$.

\begin{figure}[h] 
\centering
\scalebox{1}{\input{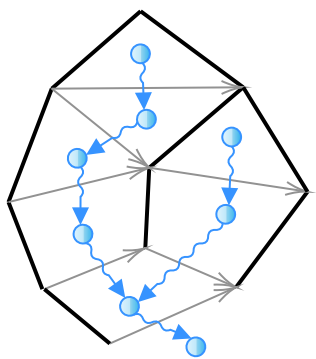}}
\caption{A graph with a spanning tree $T$ (thick black) and its corresponding interdigitating tree $T^*$ (wavy blue). $T^*$ is given a rootwise orientation, which corresponds to a counterclockwise orientation of non-tree edges in the primal (thin gray).}
\label{fig:duality}
\end{figure}

\subsection{Cut-cycle duality} A set of edges of a planar embedded
graph $G$ form a simple cycle if and only if the duals of those edges
form a simple cycle in $G^*$~\cite{whitney}.  Therefore problems
involving finding minimum-weight cuts in planar graphs are equivalent
to finding minimum-weight cycles in planar graphs.

Consider a simple cut $C=\delta(S)$ in a planar embedded graph $G$ and the
corresponding simple cycle $C^*$ in $G^*$.  The subgraph induced by
the faces of $G^*$
enclosed by $C^*$ is denoted $int(C^*)$ (the \emph{interior} of
$C^*$), and the graph induced by the faces of $G^*$
not enclosed is denoted $ext(C^*)$ (the \emph{exterior} of $C^*$).  

We define the interior and exterior of a simple cut in $G$ as follows.  The
\emph{interior} of $C$ is the planar embedded graph obtained by coalescing to a
single vertex all the vertices corresponding to the faces forming the
exterior of $C^*$.  (This is the same as taking the dual of the planar
embedded graph obtained from $G^*$ by deleting edges not enclosed by
$C^*$.)  The \emph{exterior} of $C$ is defined symmetrically.

\subsection{Chalermsook \emph{et al.}'s minimum-cut algorithm}
For planar graphs, there is an $O(n\log^2{n})$ minimum-cut algorithm
by Chalermsook, Fakcharoenphol and Nanongkai (CFN).  The algorithm
finds a simple cycle $C^*$ in the dual whose interior and exterior
have roughly the same size.  (Such a simple cycle, called a balanced
simple cycle separator, always exists if the
primal has degree three~\cite{Miller86}.)  Let $C$ be the corresponding simple cut.
The algorithm recursively finds the minimum cut in the interior of $C$
and in the exterior of $C$.  The algorithm also finds the shortest
cycle in the dual that crosses $C^*$.  This cycle corresponds in the
primal to a cut.   The minimum cut is then among
these three cuts.  The time required for this algorithm depends on how fast one can find
a shortest cycle that crosses $C^*$.  Chalermsook et al.  show that a
vertices $s$ and $t$ can be identified such that a minimum $s$-$t$ cut
in the primal corresponds in the dual to a shortest cycle cycle that
crosses $C^*$.

 We will adapt this algorithm by making two changes.  First, the
 balanced simple cycle separator is chosen to balance not the total
 graph sizes but the number of terminals.  This is straightforward.

 Second, instead of finding the shortest cycle that crosses $C^*$, the
 new algorithm must find the shortest cycle $C$ that crosses $C^*$ and
 such that $C$ encloses a face corresponding to a terminal and also
 fails to enclose some other face corresponding to a terminal.  This
 is not straightforward; we show how to modify Reif's minimum $s$-$t$
 cut algorithm to achieve this.

\subsection{Reif's minimum $s$-$t$ cut algorithm}

We now present Reif's minimum $s$-$t$ cut algorithm. Let $s^*$ and $t^*$ be the
faces of $G^*$ corresponding to vertices $s$ and $t$, respectively. Let $P^* =
f^*_1\dots f^*_l$ be a shortest path in $G^*$ between a vertex incident to $s^*$
to a vertex incident to $t^*$. Cut $G^*$ along $P^*$ by splitting each vertex $f^*$
of $P^*$ into two vertices, $f^*$ and $\hat{f}^*$, duplicating the edges on the resulting set of vertices. Considering $s^*$ as the left
end of the path and $t^*$ as the right end of the path, the edges above the
path adjacent to the original vertex are connected to $\hat{f}^*$, and the edges
below the path adjacent to the original vertex are connected to $f^*$. Note that
a shortest $\hat{f}^*$-$f^*$ path in the cut-open dual graph corresponds
directly to a cycle that passes through $f^*$ in the original dual graph.

A
shortest $s^*$-$t^*$ separating cycle must cross $P^*$. Therefore the
shortest $s^*$-$t^*$ separating cycle is equivalent to the shortest
$\hat{f}^*$-$f^*$ path, which can be found in $O(n)$ applications of Dijkstra's
algorithm. This gives an overall running time of $O(n^2\log{n})$, which was the
runtime of the original algorithm by Itai and Shiloach \cite{itai}. Reif
improved this runtime by observing the following non-crossing property.

\begin{property} \label{non-crossing}
	Let $C^*_i$ be a shortest $s^*$-$t^*$ separating cycle that crosses $f^*_i$.
Shortest separating cycles crossing vertices $f_{< i}^*$ can be found in
$int(C_i^*)$, and shortest separating cycles crossing vertices $f_{> i}^*$ can
be found in $ext(C_i^*)$. 
\end{property}

Reif's algorithm first finds the shortest cycle $C^*$ that crosses the midpoint
of $P^*$. If this is not the shortest $s^*$-$t^*$ separating cycle, then by
Property \ref{non-crossing}, a shortest $s^*$-$t^*$ separating cycle exists
entirely in $int(C^*)$ or entirely in $ext(C^*)$. Applying this algorithm
recursively on the interior and exterior subgraphs and picking the shortest
cycle yields the minimum $s$-$t$ cut. This algorithm induces a recursion tree
of depth $O(\log{n})$, where the amount of work at each level is the time it
takes to compute a shortest path on $n$ vertices, or $\textsc{SSSP}(n)$. The
algorithm thus runs in time $O(\text{SSSP}(n)\log{n})$ time. Using the
algorithm by Henzinger, Klein, Rao, and Subramanian \cite{henzinger} for the
SSSP problem in planar graphs, Reif's algorithm runs in total time
$O(n\log{n})$.

\subsection{Orientation of non-tree edges with a shortest-path tree}
Fundamentally, the CFN algorithm cannot be directly applied to the minimum
Steiner cut problem since it relies on a minimum $s$-$t$ cut algorithm as a
subroutine. Since minimum $s$-$t$ cuts may or may not be terminal-separating,
we must search for minimum terminal-separating cuts using a different method.
By choice of the terminal-separating cycle $S^*$, we can ensure that cycles
crossing $S^*$ have a terminal in their exterior. We show that a
shortest-terminal separating cycle crossing $S^*$ separates two dual faces
$v^*_e$ and $v^*$. For any dual vertex $f^*$ on the path between  $v^*_e$ and
$v^*$, a terminal-enclosing cycle can be found by orienting certain non-tree
edges with respect to a tree rooted at $f^*$ such that a cycle containing such
an edge must enclose a terminal. A shortest path computation in a layered graph
yields the shortest terminal-separating cycle. Our main insight is that we only
have to find such a cycle for one vertex on the path between $v^*_e$ and $v^*$,
since the shortest terminal-separating cycle is either this cycle or one of the
shortest cycles found by Reif's algorithm.

\section{The algorithm}
To find the minimum terminal-separating cut of a graph $G$, we find a shortest
terminal face-separating cycle in $G^*$. We first find a terminal-balancing
cycle $S^*$. Using a modified version of Reif's minimum $s$-$t$ cut algorithm,
we find the shortest terminal-separating cycle $C^*$ that crosses $S^*$. The
shortest terminal-separating cycle of $G^*$ is the smallest of: (1) The
crossing cycle $C^*$, (2) the shortest terminal-separating cycle of $int(S^*)$,
and (3) the shortest terminal-separating cycle of $ext(S^*)$. Applying the
algorithm recursively on the interior and exterior of $S^*$ yields the
solution.

\subsection{The dividing step: finding a terminal-balancing cut} \label{divide}

 Like the CFN algorithm, we need a balancing cycle separator of $G^*$. However,
for our purposes, this separator needs to balance terminals. The constructive
proof of Lemma \ref{cycle} yields such a separating cycle in linear time, after triangulating the graph.

\begin{lemma}[adaptation from \cite{klein}]\label{cycle}
	Given a triangulated graph $G^*$ with $k$ distinguished faces $X \subseteq V$
called terminals, there exists a cycle separator, which can be found in linear
time, such that the interior of the cycle contains at most 3/4 of the terminals
and the exterior contains at most 3/4 of the terminals for all $k \geq 4$.
\end{lemma}
\begin{proof}
Let $T^*$ be a spanning tree of $G^*$. Then the interdigitating tree $T$ is a
spanning tree of $G$, where some vertices are terminals. Root $T$ at a degree one
vertex $r$. For each vertex $v$, define $w(v) = \{1/k \text{ if $v$ is a
terminal}, 0 \text{ otherwise}\}$. Then define $\hat{w}(v) = \sum\big\{w(v') :
v' \text{ a descendant of } v\big\}$ and apply the function below to the root of
$T$:
\begin{equation*}
	f(v) = \begin{cases}
 		f(u), &\text{ if some child $u$ of $v$ has $\hat{w}(u) > \frac{3}{4}$} \\
 		v, &\text{ otherwise}
 	\end{cases}
\end{equation*}

\noindent Let $u = f(r)$. By induction on the number of invocations of
$f(\cdot)$, $\hat{w}(u) > \frac{3}{4}$. But $w(u) \leq \frac{1}{4}$, so $u$
must have children. Let $u'$ be the heaviest child of $u$ with respect to
$\hat{w}(\cdot)$. By choice of $u$, $\hat{w}(u') \leq \frac{3}{4}$. Since $u'$
is heaviest, $\hat{w}(u') > \frac{1}{2}(\frac{3}{4} - w(u)) \geq \frac{1}{4}$.
Let $e = uu'$. Each component of $T - \{e\}$ has total weight at most
$\frac{3}{4}$, so the simple separating cycle $S^*$ formed by $T^* \cup \{e\}$
satisfies the balance condition.

\end{proof}

We now triangulate the dual graph $G^*$ and give the new edges infinite weight to preserve the minimum Steiner cycle of the graph.
After applying Lemma \ref{cycle} to the triangulated dual graph with a
shortest path tree $T^*$ rooted at an vertex adjacent to a terminal face, we
obtain a cycle separator $S^* = P^*_a e P^*_b$, where $P^*_a$ and $P^*_b$ are
shortest paths and $e$ is the non-tree edge chosen to balance terminals.

\subsection{The conquering step: finding the minimum Steiner cut crossing $S$}
\label{conquer}
We show that there exists a shortest $S^*$-crossing cycle that separates two special faces of $G^*$. The proof is
very similar to that of Lemma 2.3 from \cite{chalermsook}. We then give a
corollary for terminal-separating cycles. See Figures \ref{fig:cfn} and
\ref{fig:cfn-terminal}.

\begin{lemma}[adaptation from \cite{chalermsook}] \label{cross}
	Let $S^* = P^*_a e P^*_b$ be a cycle separator formed by two paths of a shortest path tree $T^*$ and a non-tree edge $e$. Let $v^*_e$ be a dual face incident to $e$ and let $v^*$ be a dual face outside $S^*$ incident to the parentmost edge of $P^*_b$. Any cycle $C^*$ in $G^*$ that contains in its interior both $v^*$ and $v^*_e$ is no shorter
than a cycle that does not cross $S^*$.
\end{lemma}
\begin{proof}
Suppose $C^*$ crosses $S^*$. To contain both $v^*_e$ and $v^*$, the cycle must
cross $P^*_a$ or $P^*_b$ an even number of times. Let $f^*_{p}$ (resp.
$f^*_{l}$) be the parentmost (resp. leafmost) vertex where $C^*$ and $P^*_a$
cross with respect to $T^*$. Since $P^*_a$ is a shortest path, the subpath
$C^*[f^*_{p}, f^*_{l}]$ can be replaced with $P^*_a[f^*_{p}, f^*_{l}]$ without
increasing the length of the cycle. The subpath of $C^*$ that crosses $P^*_b$
can be replaced similarly. With these replacements, $C^*$ no longer crosses
$S^*$ and has not increased in length.
\end{proof}

\begin{corollary} \label{cross-terminal}
	Suppose $T^*$ is rooted at a vertex adjacent to a terminal face $v^*_t$. Any
terminal-separating cycle $C^*$ in $G^*$ that contains (i) both $v^*$ and
$v^*_e$ or (ii) both $v^*_t$ and $v^*_e$ is no shorter than a
terminal-separating cycle that does not cross $S^*$.
	\end{corollary}
\begin{proof}
The proof of statement (i) follows from applying the proof of Lemma \ref{cross} and noticing
that shortcutting along the paths $P^*_a$ and $P^*_b$ of $T^*$ either preserves the terminal-separating property of the cycle or not. Suppose that replacing a path $C^*[f^*, f^{*'}]$ with a shortcut $T^*[f^*, f^{*'}]$ causes the resulting cycle to no longer be terminal-separating. Then the cycle formed by the two aforementioned paths between $f^*$ and $f^{*'}$ is terminal-separating, no longer than $C^*$, and does not cross $S^*$.
Otherwise, the new cycle formed by shortcutting along $T^*$ has the properties we desire.

We now prove statement (ii). Suppose $int(C^*)$ contains $v^*_t$ and $v^*_e$.
Let $f^*_1$ be the parentmost vertex of the shortest path between $v^*_t$ and
$v^*$ that $C^*$ crosses and let $f^*_2$ and $f^*_3$ be the first and last
vertices of the subpath of $C^*$ that crosses $P^*_a$ and $P^*_b$. Since $C^*$
is terminal-separating, there is a terminal in (a) $ext(C^*) \cup ext(S^*)$ or
(b) $ext(C^*) \cup int(S^*)$. In case (a), shortcutting along the
$f^*_1$-$f^*_2$ or $f^*_1$-$f^*_3$ path in $T^*$ yields a cycle no longer than $C^*$ in
$ext(S^*)$. In case (b), shortcutting along the $f^*_1$-$f^*_2$ path and then
shortcutting from the common vertex of $P^*_a$ and $P^*_b$ to $f^*_3$ yields a
cycle no longer than $C^*$ in $int(S^*)$. If shortcutting makes the cycle no
longer terminal-separating, then there exists a subpath of $T^*$ along with a
subpath of $C^*$ which together form a terminal-separating cycle inside or
outside of $S^*$.
\end{proof}

\begin{figure}[htbp]
  \begin{minipage}[t]{0.45\linewidth}
    \centering
	\scalebox{0.8}{\tikzset{every picture/.style={line width=0.75pt}} 

\begin{tikzpicture}[x=0.75pt,y=0.75pt,yscale=-1,xscale=1]

\draw  [color={rgb, 255:red, 139; green, 139; blue, 139 }  ,draw opacity=1 ][fill={rgb, 255:red, 238; green, 238; blue, 238 }  ,fill opacity=1 ] (302.5,312) -- (344.33,337.05) -- (330.58,351.43) -- (282.42,356) -- (267.09,340.02) -- cycle ;
\draw  [color={rgb, 255:red, 139; green, 139; blue, 139 }  ,draw opacity=1 ][fill={rgb, 255:red, 238; green, 238; blue, 238 }  ,fill opacity=1 ] (330.34,114.58) -- (323.5,132) -- (284.57,107.17) -- (304.5,89) -- (319.54,95.38) -- cycle ;
\draw [line width=1.5]    (395.64,175.45) -- (401.07,217.73) ;

\draw [line width=1.5]    (204.5,275.82) -- (267.09,340.02) ;

\draw [line width=1.5]    (284.57,107.17) -- (395.64,175.45) ;

\draw [line width=1.5]    (206.65,174.34) -- (204.5,275.82) ;

\draw [line width=1.5]    (362.5,40) -- (206.65,174.34) ;

\draw [line width=1.5]    (406.5,260) -- (344.33,337.05) ;

\draw [line width=1.5]    (401.07,217.73) -- (406.5,260) ;

\draw [line width=1.5]    (362.5,40) -- (400.5,65) ;

\draw  [dash pattern={on 4.5pt off 4.5pt}] (243.5,89) .. controls (285.5,51) and (389.5,101) .. (412.5,142) .. controls (435.5,183) and (435.5,191) .. (441.5,263) .. controls (447.5,335) and (333.5,385) .. (291.5,378) .. controls (249.5,371) and (233.75,328) .. (229.5,289) .. controls (225.25,250) and (258.5,260) .. (231.5,196) .. controls (204.5,132) and (201.5,127) .. (243.5,89) -- cycle ;
\draw  [draw opacity=0][fill={rgb, 255:red, 0; green, 0; blue, 0 }  ,fill opacity=1 ] (216.33,165.17) .. controls (216.33,163.79) and (217.45,162.67) .. (218.83,162.67) .. controls (220.21,162.67) and (221.33,163.79) .. (221.33,165.17) .. controls (221.33,166.55) and (220.21,167.67) .. (218.83,167.67) .. controls (217.45,167.67) and (216.33,166.55) .. (216.33,165.17) -- cycle ;
\draw  [draw opacity=0][fill={rgb, 255:red, 0; green, 0; blue, 0 }  ,fill opacity=1 ] (228.67,302.83) .. controls (228.67,301.45) and (229.79,300.33) .. (231.17,300.33) .. controls (232.55,300.33) and (233.67,301.45) .. (233.67,302.83) .. controls (233.67,304.21) and (232.55,305.33) .. (231.17,305.33) .. controls (229.79,305.33) and (228.67,304.21) .. (228.67,302.83) -- cycle ;

\draw (345,36) node   {$T^{*}$};
\draw (309.8,103.67) node   {$v^{*}$};
\draw (303.33,332.33) node   {$v^{*}_{e}$};
\draw (232.67,166) node   {$f^{*}_{p}$};
\draw (241.33,293.33) node   {$f^{*}_{l}$};

\end{tikzpicture}}
    \caption{Proof of Lemma \ref{cross}. Thick solid tree is $T^*$ and thin
dashed cycle is $C^*$. Dual faces $v^*$ and $v^*_e$ are indicated in gray.}
    \label{fig:cfn}
  \end{minipage}
  \hspace{0.5cm}
  \begin{minipage}[t]{0.45\linewidth}
    \centering
	\scalebox{0.8}{\tikzset{every picture/.style={line width=0.75pt}} 

\begin{tikzpicture}[x=0.75pt,y=0.75pt,yscale=-1,xscale=1]

\draw  [color={rgb, 255:red, 65; green, 117; blue, 5 }  ,draw opacity=1 ][fill={rgb, 255:red, 126; green, 211; blue, 33 }  ,fill opacity=0.74 ][line width=0.75]  (369.5,60) -- (342.5,74) -- (321.5,56) -- (335.5,40) -- (359.5,39) -- cycle ;
\draw  [color={rgb, 255:red, 139; green, 139; blue, 139 }  ,draw opacity=1 ][fill={rgb, 255:red, 238; green, 238; blue, 238 }  ,fill opacity=1 ] (261.5,328) -- (303.33,353.05) -- (289.58,367.43) -- (241.42,372) -- (226.09,356.02) -- cycle ;
\draw  [color={rgb, 255:red, 139; green, 139; blue, 139 }  ,draw opacity=1 ][fill={rgb, 255:red, 238; green, 238; blue, 238 }  ,fill opacity=1 ] (289.34,130.58) -- (282.5,148) -- (243.57,123.17) -- (263.5,105) -- (278.54,111.38) -- cycle ;
\draw [line width=1.5]    (354.64,191.45) -- (360.07,233.73) ;

\draw [line width=1.5]    (163.5,291.82) -- (226.09,356.02) ;

\draw [line width=1.5]    (243.57,123.17) -- (354.64,191.45) ;

\draw [line width=1.5]    (165.65,190.34) -- (163.5,291.82) ;

\draw [line width=1.5]    (321.5,56) -- (165.65,190.34) ;

\draw [line width=1.5]    (365.5,276) -- (303.33,353.05) ;

\draw [line width=1.5]    (360.07,233.73) -- (365.5,276) ;

\draw  [dash pattern={on 4.5pt off 4.5pt}] (291.5,27) .. controls (333.5,-11) and (373.5,21) .. (396.5,62) .. controls (419.5,103) and (400.5,215) .. (406.5,287) .. controls (412.5,359) and (292.5,401) .. (250.5,394) .. controls (208.5,387) and (192.75,344) .. (188.5,305) .. controls (184.25,266) and (341.5,197) .. (314.5,133) .. controls (287.5,69) and (249.5,65) .. (291.5,27) -- cycle ;
\draw  [draw opacity=0][fill={rgb, 255:red, 0; green, 0; blue, 0 }  ,fill opacity=1 ] (284.33,86.17) .. controls (284.33,84.79) and (285.45,83.67) .. (286.83,83.67) .. controls (288.21,83.67) and (289.33,84.79) .. (289.33,86.17) .. controls (289.33,87.55) and (288.21,88.67) .. (286.83,88.67) .. controls (285.45,88.67) and (284.33,87.55) .. (284.33,86.17) -- cycle ;
\draw  [draw opacity=0][fill={rgb, 255:red, 0; green, 0; blue, 0 }  ,fill opacity=1 ] (187.67,318.83) .. controls (187.67,317.45) and (188.79,316.33) .. (190.17,316.33) .. controls (191.55,316.33) and (192.67,317.45) .. (192.67,318.83) .. controls (192.67,320.21) and (191.55,321.33) .. (190.17,321.33) .. controls (188.79,321.33) and (187.67,320.21) .. (187.67,318.83) -- cycle ;
\draw  [color={rgb, 255:red, 65; green, 117; blue, 5 }  ,draw opacity=1 ][fill={rgb, 255:red, 188; green, 224; blue, 149 }  ,fill opacity=0.74 ][dash pattern={on 0.84pt off 2.51pt}][line width=0.75]  (245.5,191) -- (216.5,200) -- (212.5,184) -- (224.5,167) -- (244.5,170) -- cycle ;
\draw  [color={rgb, 255:red, 65; green, 117; blue, 5 }  ,draw opacity=1 ][fill={rgb, 255:red, 188; green, 224; blue, 149 }  ,fill opacity=0.74 ][dash pattern={on 0.84pt off 2.51pt}][line width=0.75]  (215.5,100) -- (192.5,104) -- (180.5,91) -- (194.5,75) -- (214.5,79) -- cycle ;
\draw  [draw opacity=0][fill={rgb, 255:red, 0; green, 0; blue, 0 }  ,fill opacity=1 ] (311.33,166.17) .. controls (311.33,164.79) and (312.45,163.67) .. (313.83,163.67) .. controls (315.21,163.67) and (316.33,164.79) .. (316.33,166.17) .. controls (316.33,167.55) and (315.21,168.67) .. (313.83,168.67) .. controls (312.45,168.67) and (311.33,167.55) .. (311.33,166.17) -- cycle ;

\draw (269.8,119.67) node   {$v^{*}$};
\draw (262.33,348.33) node   {$v^{*}_{e}$};
\draw (265.67,79) node   {$f^{*}_{1}$};
\draw (202.33,306.33) node   {$f^{*}_{3}$};
\draw (345.8,52.67) node   {$v^{*}_{t}$};
\draw (231.8,211.67) node   {$( b)$};
\draw (201.8,116.67) node   {$( a)$};
\draw (331.67,157) node   {$f^{*}_{2}$};

\end{tikzpicture}}
    \caption{Proof of Corollary \ref{cross-terminal}. Light green faces are
terminals in (a) $ext(C^*) \cup ext(S^*)$ or in (b) $ext(C^*) \cup int(S^*)$}
    \label{fig:cfn-terminal}
  \end{minipage}
\end{figure}

Therefore, to find a shortest terminal-separating cycle crossing $S^*$, it suffices to find a shortest terminal-separating cycle $C^*$
that separates $v^*_t$ and $v^*$ from $v^*_e$. In the same
manner as Reif's algorithm, we create a new graph $\hat{G}^*$ by cutting along
a path. To ensure that $v^*_t$ is in $ext(C^*)$, we cut along the path $P^*_a$
extended to the root. Recall that a shortest path in $\hat{G}^*$ between a vertex
$f^*$ of $P^*_a$ and its copy $\hat{f}^*$ corresponds directly to a shortest
$v^*_e$-$v^*$ separating cycle crossing the vertex $f^*$ in $G^*$.

We now use a slightly modified version of Reif's algorithm to produce 
$v^*_e$-$v^*$ separating cycles in $G^*$ and show that either the shortest
terminal-separating cycle is found by Reif's algorithm, or it can be found with
Dijkstra's algorithm on a new graph construction. We modify Reif's algorithm by
observing a non-crossing property which is an extension of Property
\ref{non-crossing}.

\begin{property} \label{non-crossing-terminal}
	Let $C^*$ be a shortest $v^*_e$-$v^*$ separating cycle that does not separate
terminals. There exists a shortest $v^*_e$-$v^*$ separating,
terminal-separating cycle that does not cross $C^*$.
\end{property}
\begin{proof}
	Suppose without loss of generality that $ext(C^*)$ contains all of the terminals. Let $\tilde{C}^*$ be a $v^*_e$-$v^*$ separating, terminal-separating cycle that crosses $C^*$. $\tilde{C}^*$ must cross $C^*$ an even number of times. We will now construct a cycle as short as $\tilde{C}^*$ contained in $ext(C^*)$. For every consecutive pair of vertices where the two cycles cross, consider the two possible paths. If the path along $\tilde{C}^*$ is enclosed by $int(C^*)$, take the path along $C^*$. Otherwise, take the path along $\tilde{C^*}$. The resulting cycle is terminal-separating since $int(C^*)$ contains no terminals, and the choice of paths ensures that it is in $ext(C^*)$.
\end{proof}

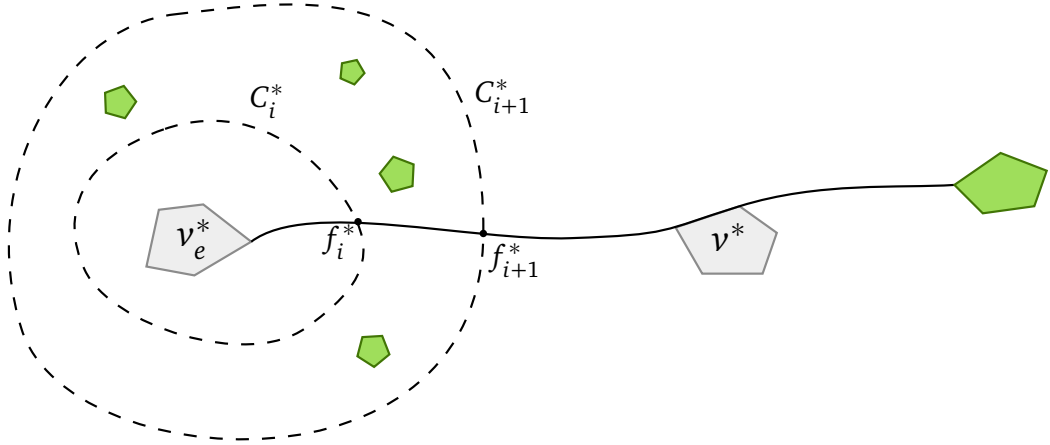
\begin{figure}[h] 
\centering
\scalebox{1.1}{\tikzset{every picture/.style={line width=0.75pt}} 

\begin{tikzpicture}[x=0.75pt,y=0.75pt,yscale=-0.75,xscale=0.75]

\draw  [color={rgb, 255:red, 139; green, 139; blue, 139 }  ,draw opacity=1 ][fill={rgb, 255:red, 238; green, 238; blue, 238 }  ,fill opacity=1 ] (132.87,194.64) -- (162.18,217.44) -- (127.44,238.06) -- (98.13,232.63) -- (105.73,197.9) -- cycle ;
\draw  [color={rgb, 255:red, 139; green, 139; blue, 139 }  ,draw opacity=1 ][fill={rgb, 255:red, 238; green, 238; blue, 238 }  ,fill opacity=1 ] (474.81,236.98) -- (437.9,236.98) -- (421.62,208.75) -- (460.7,195.73) -- (483.5,212.01) -- cycle ;
\draw    (162.18,217.44) .. controls (197.23,191.14) and (295.7,217.44) .. (357.57,215.27) .. controls (419.45,213.09) and (411.85,210.92) .. (460.7,195.73) .. controls (509.55,180.53) and (558.4,184.87) .. (592.05,182.7) ;

\draw  [dash pattern={on 4.5pt off 4.5pt}] (109.51,148.42) .. controls (161.22,130.89) and (201.54,163.32) .. (220.82,194) .. controls (240.11,224.68) and (231.34,244.83) .. (195.41,270.25) .. controls (159.47,295.67) and (76.21,268.5) .. (60.43,234.75) .. controls (44.65,201.01) and (57.8,165.95) .. (109.51,148.42) -- cycle ;
\draw  [dash pattern={on 4.5pt off 4.5pt}] (115.65,78.3) .. controls (186.64,70.42) and (304.96,45.87) .. (304.09,212.4) .. controls (303.21,378.93) and (49.91,360.53) .. (21.86,271.13) .. controls (-6.18,181.73) and (44.65,86.19) .. (115.65,78.3) -- cycle ;
\draw  [draw opacity=0][fill={rgb, 255:red, 0; green, 0; blue, 0 }  ,fill opacity=1 ] (225.23,205.21) .. controls (225.23,203.96) and (226.24,202.95) .. (227.49,202.95) .. controls (228.74,202.95) and (229.75,203.96) .. (229.75,205.21) .. controls (229.75,206.46) and (228.74,207.48) .. (227.49,207.48) .. controls (226.24,207.48) and (225.23,206.46) .. (225.23,205.21) -- cycle ;
\draw  [draw opacity=0][fill={rgb, 255:red, 0; green, 0; blue, 0 }  ,fill opacity=1 ] (301.82,212.4) .. controls (301.82,211.15) and (302.84,210.14) .. (304.09,210.14) .. controls (305.34,210.14) and (306.35,211.15) .. (306.35,212.4) .. controls (306.35,213.66) and (305.34,214.67) .. (304.09,214.67) .. controls (302.84,214.67) and (301.82,213.66) .. (301.82,212.4) -- cycle ;
\draw  [color={rgb, 255:red, 65; green, 117; blue, 5 }  ,draw opacity=1 ][fill={rgb, 255:red, 126; green, 211; blue, 33 }  ,fill opacity=0.74 ] (226.75,106.29) -- (231.42,114.3) -- (225.24,121.22) -- (216.76,117.48) -- (217.69,108.26) -- cycle ;
\draw  [color={rgb, 255:red, 65; green, 117; blue, 5 }  ,draw opacity=1 ][fill={rgb, 255:red, 126; green, 211; blue, 33 }  ,fill opacity=0.74 ] (85.05,141.87) -- (73.31,138.55) -- (72.84,126.35) -- (84.3,122.14) -- (91.84,131.73) -- cycle ;
\draw  [color={rgb, 255:red, 65; green, 117; blue, 5 }  ,draw opacity=1 ][fill={rgb, 255:red, 126; green, 211; blue, 33 }  ,fill opacity=0.74 ] (260.8,183.41) -- (248,186.87) -- (240.75,175.77) -- (249.07,165.44) -- (261.46,170.17) -- cycle ;
\draw  [color={rgb, 255:red, 65; green, 117; blue, 5 }  ,draw opacity=1 ][fill={rgb, 255:red, 126; green, 211; blue, 33 }  ,fill opacity=0.74 ] (246.66,286.37) -- (237.3,294.05) -- (227.1,287.52) -- (230.16,275.8) -- (242.25,275.09) -- cycle ;
\draw  [color={rgb, 255:red, 65; green, 117; blue, 5 }  ,draw opacity=1 ][fill={rgb, 255:red, 126; green, 211; blue, 33 }  ,fill opacity=0.74 ] (640.9,195.73) -- (609.42,200.07) -- (592.05,182.7) -- (620.28,163.16) -- (648.5,174.02) -- cycle ;

\draw (127.02,215.79) node [scale=1.2]  {$v^{*}_{e}$};
\draw (453.91,214.74) node [scale=1.2]  {$v^{*}$};
\draw (216.09,217.37) node   {$f^{*}_{i}$};
\draw (323.56,228.77) node   {$f^{*}_{i+1}$};
\draw (171.58,131.61) node   {$C^{*}_{i}$};
\draw (315.96,128.36) node   {$C^{*}_{i+1}$};

\end{tikzpicture}}
\caption{The cycles $C^*_i$ and $C^*_{i+1}$ found by Reif's algorithm. Green
faces are terminals.}
\label{fig:reif}
\end{figure}

We can therefore modify Reif's algorithm to stop recursing on the interior of a
cycle if the interior does not contain terminals. Recall that the path between
$v^*_e$ and $v^*$ is $P^*_a = f^*_1\dots f^*_l$. Let the shortest cycles
crossing each vertex of $P^*_a$ found by Reif's algorithm be $C^*_1 \dots
C^*_l$. Let $i = \max \big\{ k = 1 \dots l \mid int(C^*_k) \text{ contains no
terminals} \big\}$. If $i = l$, there exists a subpath of $C^*_l$ enclosed
by $S^*$ which forms a terminal-separating cycle when connected to a subpath
of $T^*$ since $int(S^*)$ contains terminals. This cycle is no longer than
 $C^*_l$ and is contained in $int(S^*)$. Now suppose $i < l$.
 By choice of $i$, the cycles $C^*_{i+1}, C^*_{i+2}, \dots, C^*_{l}$
 are terminal-separating. The shortest
 terminal-separating cycle is either one of these, or by the non-crossing
properties it is the shortest-terminal separating cycle crossing $f^*_i$ in the
annulus graph $G^*_i = ext(C^*_i) \cap int(C^*_{i+1})$.

It now remains to find the shortest terminal-separating cycle crossing $f^*_i$
in $G^*_i$. Let $T^*_i$ be a shortest-path tree rooted at $f^*_i$ in $G^*_i$
and let $T_i$ be its interdigitating tree. The next lemma allows us to
restrict our focus to cycles that enclose the fundamental cycle of each of its
non-tree edges. 

\begin{lemma} \label{fundamental-cycles}
There exists a shortest terminal-enclosing cycle $\tilde{C}^*_i$ crossing
$f^*_i$ such that, for every non-tree edge $e$, the fundamental cycle of $e$ in
$T^*_i$ is contained in
$int(\tilde{C}^*_i)$. 
\end{lemma}

\begin{proof}
Let $\tilde{C}^*_i$ be a shortest terminal-enclosing cycle crossing $f^*_i$.
Consider the leafmost edges of $T_i$ such that their fundamental cycles in
$T^*_i$ are not enclosed by $\tilde{C}^*_i$. Combining these cycles with
$\tilde{C}^*_i$ gives a cycle no longer than $\tilde{C}^*_i$ since each
disjoint fundamental cycle replaces some subpath of $\tilde{C}^*_i$ with a
shortest path, and the combination of adjacent fundamental cycles also replaces
some subpath of $\tilde{C}^*_i$ with a shortest path. Repeating this process on
$T_i$ results in a terminal-enclosing cycle no shorter than $\tilde{C}^*_i$
crossing $f^*_i$. 
\end{proof}

\begin{figure}[h] 
\centering
\scalebox{1}{\tikzset{every picture/.style={line width=0.75pt}} 

\begin{tikzpicture}[x=0.75pt,y=0.75pt,yscale=-1,xscale=1]
height of 432

\draw [line width=1.5]    (242.55,145.88) -- (163.5,261.16) ;

\draw [line width=1.5]    (290.92,196.32) -- (213.39,303.63) ;

\draw [line width=1.5]    (339.29,246.75) -- (242.55,145.88) ;

\draw [line width=1.5]    (272.84,339.72) -- (339.29,246.75) ;

\draw [line width=1.5]    (366.26,333.35) -- (339.29,246.75) ;

\draw [line width=1.5]    (274.96,95.56) -- (242.55,145.88) ;

\draw  [dash pattern={on 4.5pt off 4.5pt}]  (280.27,96.62) .. controls
(302.56,64.78) and (367.32,109.36) .. (390.67,155.01) .. controls
(414.03,200.66) and (449.06,305.75) .. (373.69,350.34) .. controls
(298.32,394.92) and (242.23,374.04) .. (281.51,334.77) ;

\draw [color={rgb, 255:red, 74; green, 144; blue, 226 }  ,draw opacity=1 ][line
width=1.5]    (363.6,307.87) -- (284.36,333.83) ;
\draw [shift={(281.51,334.77)}, rotate = 341.86] [color={rgb, 255:red, 74;
green, 144; blue, 226 }  ,draw opacity=1 ][line width=1.5]    (14.21,-4.28) ..
controls (9.04,-1.82) and (4.3,-0.39) .. (0,0) .. controls (4.3,0.39) and
(9.04,1.82) .. (14.21,4.28)   ;

\draw  [dash pattern={on 4.5pt off 4.5pt}]  (363.6,307.87) .. controls
(358.03,294.43) and (347.24,252.5) .. (343.52,245.6) .. controls (339.81,238.7)
and (310.35,210.74) .. (296.81,195.7) ;

\draw  [dash pattern={on 4.5pt off 4.5pt}]  (296.81,195.7) -- (253.91,254.97) ;

\draw [color={rgb, 255:red, 208; green, 2; blue, 27 }  ,draw opacity=1 ][line
width=1.5]    (182.7,239.76) -- (250.98,254.35) ;
\draw [shift={(253.91,254.97)}, rotate = 192.06] [color={rgb, 255:red, 208;
green, 2; blue, 27 }  ,draw opacity=1 ][line width=1.5]    (14.21,-4.28) ..
controls (9.04,-1.82) and (4.3,-0.39) .. (0,0) .. controls (4.3,0.39) and
(9.04,1.82) .. (14.21,4.28)   ;

\draw  [dash pattern={on 4.5pt off 4.5pt}]  (280.27,96.62) -- (182.7,239.76) ;

\draw (327,304) node   {$d^{*}_{\circlearrowright }$};
\draw (220,231) node   {$d^{*}_{\circlearrowleft }$};

\end{tikzpicture}}
\caption{Example of a simple cycle crossing the root of $T^*_i$ which contains
a counterclockwise dart $d^*_\circlearrowleft$ (shown in red) and a
clockwise dart $d^*_\circlearrowright$ (shown in blue). }
\label{fig:crossing-proof}
\end{figure}
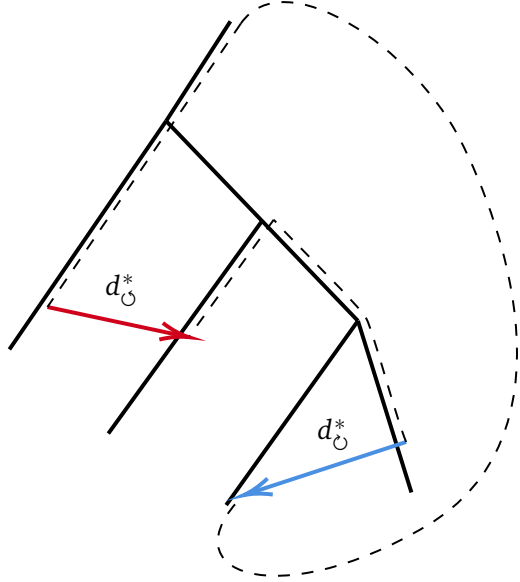

We now classify the edges that are not in $T^*_i$ to find
terminal-enclosing cycles. Consider the
counterclockwise orientation $\lambda_{T_i}: E[T_i] \rightarrow D[T_i]$ of
non-tree edges (with respect to $T^*_i$). Recall that this corresponds to the rootwise orientation of edges in $T_i$. In the following lemma, we show that a
simple counterclockwise cycle must agree with $\lambda_{T_i}$ to enclose its
fundamental cycles. Refer to Figure \ref{fig:crossing-proof}.

\begin{lemma} \label{counterclockwise}
	A simple counterclockwise cycle $C^*$ crossing the root of $T^*_i$ encloses
the fundamental cycles of its non-tree edges if and only if it agrees with
$\lambda_{T_i}$.
\end{lemma}
\begin{proof}
Suppose that $C^*$ does not agree with $\lambda_{T_i}$. Without loss of generality suppose $C^*$ contains a counterclockwise non-tree
dart $d^*_\circlearrowleft$ and a clockwise non-tree dart
$d^*_\circlearrowright$. The fundamental cycle of either $d^*_\circlearrowleft$
or $d^*_\circlearrowright$ in $T^*_i$ is not enclosed by $C^*$, otherwise $C^*$
is non-simple. On the other hand, suppose that $C^*$ does not enclose the fundamental cycle of one of its edges $e$. Consider the subtree of the interdigitating tree $T_i$ that is rooted at the parentmost vertex of $e$. $C^*$ must contain another edge $e'$ of this subtree since it does not enclose the fundamental cycle of $e$. But since $C^*$ is simple, it must contain the clockwise dart of $e$ or $e'$. So $C^*$ does not agree with $\lambda_{T_i}$.
\end{proof}

By Lemmas \ref{fundamental-cycles} and \ref{counterclockwise}, we can limit our search to cycles that agree with $\lambda_{T_i}$ without sacrificing optimality. We label the interior and exterior faces of the annulus graph as $v^*_{int}$
and $v^*_{ext}$, respectively. Root the interdigitating tree $T_i$ at
$v_{ext}$. Let $R = \{  u \rightarrow v \in \lambda_{T_i} \mid uv \text{ is an
ancestor of a terminal in } T_i \}$ be a set of counterclockwise darts which we
color \emph{red}. Treating $v^*_{ext}$ as the infinite face, every edge of
$T_i$ that is a parent of a terminal vertex corresponds to a fundamental cycle in the dual
enclosing the terminal face. By the following lemma, the presence of red darts sufficiently
characterizes terminal-enclosing cycles crossing $f^*_i$.

\begin{lemma} \label{red-edge}
Let $\tilde{C}^*_i$ be a simple counterclockwise $v^*_{int}$-$v^*_{ext}$ separating cycle in $G^*_i$ that crosses
$f^*_i$ and agrees with $\lambda_{T_i}$. $\tilde{C}^*_i$ encloses a terminal if and only if it has a red dart.
\end{lemma}

\begin{proof}
Suppose $\tilde{C}^*_i$ does not have a red
dart. $\tilde{C}^*_i$ must contain at least one non-tree dart. But the only non-tree darts left are counterclockwise darts that do not enclose terminals or clockwise non-tree darts. $\tilde{C}^*_i$ cannot contain clockwise non-tree darts since it agrees with $\lambda_{T_i}$. So $\tilde{C}^*_i$ does not enclose a terminal. On the other hand, suppose $\tilde{C}^*_i$ has a red dart. Since this dart is a parent of a terminal with respect to $T_i$, it corresponds to a fundamental cycle enclosing a terminal. Then by Lemmas
\ref{fundamental-cycles} and \ref{counterclockwise}, $int(\tilde{C}^*_i)$
contains a terminal. \end{proof}

\begin{figure}[h] 
\centering
\scalebox{1}{\tikzset{every picture/.style={line width=0.75pt}} 

\begin{tikzpicture}[x=0.75pt,y=0.75pt,yscale=-1,xscale=1]

\draw  [fill={rgb, 255:red, 255; green, 255; blue, 255 }  ,fill opacity=0.8 ] (315.83,141.09) -- (460.5,141.09) -- (427.12,252) -- (282.45,252) -- cycle ;
\draw [line width=1.5]    (360.69,238.18) .. controls (371.1,239.27) and (388.98,231.09) .. (401.65,216.55) .. controls (414.31,202) and (407.61,166.55) .. (383.06,173.66) ;

\draw  [fill={rgb, 255:red, 255; green, 255; blue, 255 }  ,fill opacity=1 ] (380.29,173.66) .. controls (380.29,171.79) and (381.53,170.28) .. (383.06,170.28) .. controls (384.59,170.28) and (385.82,171.79) .. (385.82,173.66) .. controls (385.82,175.52) and (384.59,177.04) .. (383.06,177.04) .. controls (381.53,177.04) and (380.29,175.52) .. (380.29,173.66) -- cycle ;
\draw  [fill={rgb, 255:red, 255; green, 255; blue, 255 }  ,fill opacity=0.8 ] (302.42,72) -- (447.09,72) -- (413.71,182.91) -- (269.04,182.91) -- cycle ;
\draw  [fill={rgb, 255:red, 255; green, 255; blue, 255 }  ,fill opacity=0.8 ] (67.88,117.45) -- (212.55,117.45) -- (179.17,228.36) -- (34.5,228.36) -- cycle ;
\draw [color={rgb, 255:red, 208; green, 2; blue, 27 }  ,draw opacity=1 ][line width=1.5]    (85.39,213.02) -- (109.75,213.57) ;
\draw [shift={(112.74,213.63)}, rotate = 181.29] [color={rgb, 255:red, 208; green, 2; blue, 27 }  ,draw opacity=1 ][line width=1.5]    (14.21,-4.28) .. controls (9.04,-1.82) and (4.3,-0.39) .. (0,0) .. controls (4.3,0.39) and (9.04,1.82) .. (14.21,4.28)   ;

\draw [line width=1.5]    (113.25,145.43) .. controls (85.56,144.82) and (62.06,163.82) .. (78.45,176.55) .. controls (94.84,189.27) and (50.89,209.27) .. (85.39,213.02) ;

\draw [line width=1.5]    (112.74,213.63) .. controls (123.15,214.73) and (141.03,206.55) .. (153.7,192) .. controls (166.36,177.45) and (159.66,142) .. (135.11,149.11) ;

\draw  [fill={rgb, 255:red, 255; green, 255; blue, 255 }  ,fill opacity=1 ] (113.25,145.43) .. controls (113.25,143.57) and (114.49,142.05) .. (116.02,142.05) .. controls (117.55,142.05) and (118.79,143.57) .. (118.79,145.43) .. controls (118.79,147.3) and (117.55,148.81) .. (116.02,148.81) .. controls (114.49,148.81) and (113.25,147.3) .. (113.25,145.43) -- cycle ;
\draw  [fill={rgb, 255:red, 255; green, 255; blue, 255 }  ,fill opacity=1 ] (132.34,149.11) .. controls (132.34,147.25) and (133.58,145.73) .. (135.11,145.73) .. controls (136.64,145.73) and (137.88,147.25) .. (137.88,149.11) .. controls (137.88,150.98) and (136.64,152.49) .. (135.11,152.49) .. controls (133.58,152.49) and (132.34,150.98) .. (132.34,149.11) -- cycle ;
\draw [color={rgb, 255:red, 208; green, 2; blue, 27 }  ,draw opacity=1 ][line width=1.5]    (321.42,169.38) -- (359.2,235.57) ;
\draw [shift={(360.69,238.18)}, rotate = 240.28] [color={rgb, 255:red, 208; green, 2; blue, 27 }  ,draw opacity=1 ][line width=1.5]    (14.21,-4.28) .. controls (9.04,-1.82) and (4.3,-0.39) .. (0,0) .. controls (4.3,0.39) and (9.04,1.82) .. (14.21,4.28)   ;

\draw [line width=1.5]    (349.28,101.8) .. controls (321.58,101.18) and (298.09,120.18) .. (314.48,132.91) .. controls (330.87,145.64) and (286.92,165.64) .. (321.42,169.38) ;

\draw  [fill={rgb, 255:red, 255; green, 255; blue, 255 }  ,fill opacity=1 ] (349.28,101.8) .. controls (349.28,99.93) and (350.52,98.42) .. (352.04,98.42) .. controls (353.57,98.42) and (354.81,99.93) .. (354.81,101.8) .. controls (354.81,103.66) and (353.57,105.18) .. (352.04,105.18) .. controls (350.52,105.18) and (349.28,103.66) .. (349.28,101.8) -- cycle ;
\draw [line width=1.5]    (216.24,165.64) -- (252.35,165.64) ;
\draw [shift={(255.35,165.64)}, rotate = 180] [color={rgb, 255:red, 0; green, 0; blue, 0 }  ][line width=1.5]    (14.21,-4.28) .. controls (9.04,-1.82) and (4.3,-0.39) .. (0,0) .. controls (4.3,0.39) and (9.04,1.82) .. (14.21,4.28)   ;

\draw (119.06,130.18) node  [font=\footnotesize]  {$f^{*}_{i}$};
\draw (138.43,131.09) node  [font=\footnotesize]  {$\hat{f_{i}}^{*}$};
\draw (379.75,117.85) node    {$\hat{G}^{*}_{i}$};
\draw (350.87,85.45) node  [font=\footnotesize]  {$f^{*}_{i}$};
\draw (374.52,204.76) node    {$\hat{G}^{*'}_{i}$};
\draw (95.28,199.3) node    {$uv$};
\draw (329.07,159.3) node    {$u$};
\draw (349.19,241.21) node    {$v'$};
\draw (372.43,166.09) node  {$\textcolor{gray}{\hat{f_{i}}^{*'}}$};

\end{tikzpicture}}
\caption{Graph construction used to find a shortest terminal separating cycle in $G^*_i$. The graph on the left is $G^*_i$ split along the edge $f^*_i f^*_{i+1}$, and the graph on the right is the layered graph used to find terminal separating cycles. }
\label{fig:graph-construction}
\end{figure}
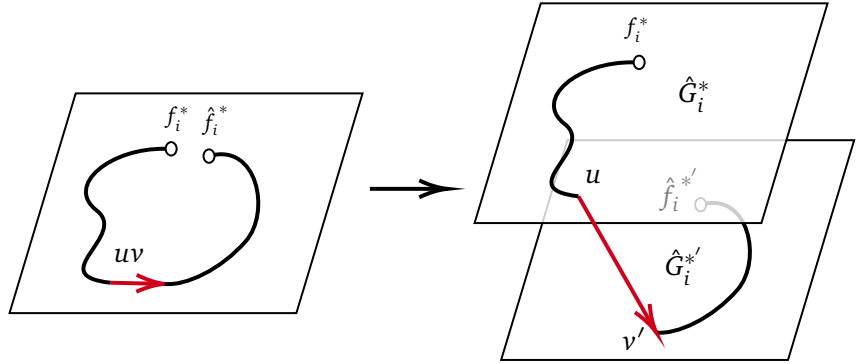

We now construct a graph that can be used to find a shortest terminal separating cycle in $G^*_i$ (refer to Figure \ref{fig:graph-construction}). First split $G^*_i$ along the edge $f^*_i f^*_{i+1}$,
resulting in a copied edge $\hat{f}^*_i \hat{f}^*_{i+1}$. Then replace each edge $uv \in E[G^*_i]$ with two directed edges $u \rightarrow v$ and $v \rightarrow u$. Call this new graph $\hat{G}^*_i$.
Let $\hat{G}^{*'}_i$ be a copy of $\hat{G}^*_i$. For each dart $u
\rightarrow v \in R$, replace $u \rightarrow v$ with an edge $u \rightarrow v'$ where $v'$ is $\hat{G}^{*'}_i$'s copy of $v$. Remove each edge in $\hat{G}^*_i$ and $\hat{G}^{*'}_i$ that does not agree with $\lambda_{T_i}$ (by Lemmas \ref{fundamental-cycles} and \ref{counterclockwise} this will not limit our solution).
A shortest path in this new layered graph from $f^*_i \in \hat{G}^*_i$ to
$\hat{f}^{*'}_i \in \hat{G}^{*'}_i$ is equivalent to a shortest counterclockwise cycle
crossing $f^*_i$ that has a red dart and agrees with $\lambda_{T_i}$, which is a shortest terminal-separating
cycle crossing $f^*_i$. Running Dijkstra's algorithm from $f^*_i$ to $\hat{f}^{*'}_i$ gives
such a cycle.

\section{Time complexity}

We break up the complexity into the dividing step and the conquering step. In
the dividing step, the operations for finding a cycle separator and creating
subgraphs both take linear time. In the conquering step, we first modify Reif's
algorithm by checking the interior of cycles, which adds at most $O(n)$ time to
any level of the recursion tree. Hence, the modified Reif's algorithm still
runs in $O(n\log{n})$ time. The darts can be colored in linear time with a traversal of $T_i$. The construction of the layered graph takes
linear time and the additional call to Dijkstra's algorithm takes $O(|E| +
|V|\log{|V|})$ time, which is $O(n\log{n})$ since $|E| = O(|V|)$ in the layered
graph. So the conquering step takes a total of $O(n\log{n})$ time.

Since the divide step decreases the number of terminals geometrically, the
recursion tree of the entire algorithm has $O(\log{k})$ depth, where $k$ is the
number of terminals. Every level of this tree takes $O(n\log{n})$ time, so the
entire algorithm runs in $O(n\log{n}\log{k})$ time, proving Theorem
\ref{algorithm}.

\section{Remarks}
We have presented an algorithm for finding minimum Steiner cuts in near-linear
time in planar graphs. Since our algorithm relies on the characterization of
individual edges and faces of the dual graph, it is not trivial to adapt it to
a dense-distance graph based algorithm such as that of Italiano \emph{et al.}
\cite{italiano} or Łącki \emph{et al.} \cite{lacki}. However, we conjecture
that the runtime could be improved to at least $O(n\log\log{n}\log{k})$ with
clever use of similar techniques.

\bibliographystyle{plain}
\bibliography{terminal-cut-arxiv-combined}

\end{document}